\documentclass[12pt]{./iopart}
\expandafter\let\csname equation*\endcsname\relax
\expandafter\let\csname endequation*\endcsname\relax
\usepackage{IEEEtrantools}
\usepackage{fullpage}

\usepackage{cite}
\usepackage{amsmath}
\usepackage{amssymb}
\usepackage{amsthm}
\usepackage{graphicx}
\usepackage{color}

\theoremstyle{definition}

\theoremstyle{remark}

\newcommand{\vm}[1]{\boldsymbol{#1}}

\newcommand{\M}{M}
\newcommand{\N}{N}

\newcommand{\nb}{T}
\newcommand{\Q}{{\rm Q}}
\newcommand{\G}{{\rm G}}

\begin{document}

\title{Typical $l_1$-recovery limit of sparse vectors represented by concatenations of random orthogonal matrices}

\author{$^\dag$Yoshiyuki Kabashima, $^{\ddag,\S}$Mikko Vehkaper\"{a} and
$^{\ddag}$Saikat Chatterjee}

\address{$^\dag$Department of Computational Intelligence and Systems Science, 
Tokyo Institute of Technology, 
Yokohama 226-8502, Japan \\
$^\ddag$KTH Royal Institute of Technology and the ACCESS Linnaeus Center, SE-10044, Stockholm, Sweden,\\
$^\S$Aalto University, School of Electrical Engineering, P.O. Box 13000,
FI-00076 Aalto, Finland}
\ead{$^\dag$kaba@dis.titech.ac.jp, $^\S$mikko.vehkapera@aalto.fi, $^\ddag$sach@kth.se}


\begin{abstract}
We consider the problem of recovering an $N$-dimensional sparse vector $\vm{x}$ from 
its linear transformation $\vm{y}=\vm{D} \vm{x}$ of $M(< N)$ dimension. Minimizing the $l_{1}$-norm 
of $\vm{x}$ under the constraint $\vm{y} = \vm{D} \vm{x}$ is a standard approach for 
the recovery problem, and earlier studies report that the critical condition for 
typically successful $l_1$-recovery is universal over a variety of randomly constructed 
matrices $\vm{D}$. For examining the extent of the universality, we focus on the case in 
which $\vm{D}$ is provided by concatenating $\nb=N/M$ matrices 
$\vm{O}_{1}, \vm{O}_{2}, \ldots, \vm{O}_\nb$ drawn uniformly according to the Haar measure 
on the $M \times M$ orthogonal matrices. 
By using the replica method in conjunction with the development of 
an integral formula for handling the random orthogonal matrices, we show that the
concatenated matrices can result in better recovery performance than what  
the universality predicts when the density of non-zero signals is not uniform among the
$\nb$ matrix modules. The universal condition is reproduced for the special case of
uniform non-zero signal densities.  Extensive numerical experiments support the
theoretical predictions.  
\end{abstract}

\maketitle

\section{Introduction}
The recovery problem of sparse vectors from a linear underdetermined set of equations has recently attracted attention in various fields of science and technology due to its many applications, for example, in linear regression \cite{Miller_2002_Subset_Selection_in_Regression}, communication \cite{Fletcher_ISIT_2009}, \cite{Bajwa_2010_Sparse_Multipath_Channels}, \cite{Gastpar_Sastry_2010_Distributed_sensor_perception}, multimedia \cite{Plumbley_Proc_IEEE_2010}, \cite{Sastry_2009_Face_recognition}, \cite{Elad_book_2010}, and  compressive sampling (CS) \cite{Donoho_2006_Compressed_sensing}, \cite{CS_introduction_Candes_Wakin_2008}. In such a sparse representation problem, we have the following underdetermined set of linear equations
\begin{equation}
\vm{y} = \vm{D} \vm{x},
\label{eq:Sparse_Representation_without_Noise}
\end{equation}
where $\vm{y} \in \mathbb{R}^{\M}$ is 
\textcolor{black}{a vector of interest,}  
$\vm{D} \in \mathbb{R}^{\M \times \N}$ is the dictionary 
\textcolor{black}{(sparsity inducing basis),} 
$\vm{x} \in \mathbb{R}^{\N}$ is 
\textcolor{black}{
the sparse expression of $\vm{y}$,}
and $\M < \N$
\footnote{\textcolor{black}{In a simple setup of CS, $\vm{D}$ is handled as a sensing matrix that can be 
designed. Recent studies show that the signal (vector) recovery performance can be improved significantly 
by optimizing the design of $\vm{D}$ in conjunction with the use of an approximate Bayesian inference scheme \cite{Krzakala2012a,Krzakala2012b}. 
However, an appropriate dictionary that induces sparse expressions is generally determined by the nature of the signal of interest, 
and therefore, cannot be designed freely. The purpose of this paper is to investigate a sparsity inducing property 
of such a predetermined dictionary when the widely used $l_1$-recovery scheme is employed. 
}}.
Another way of 
writing \eqref{eq:Sparse_Representation_without_Noise} is that a large dimensional 
sparse vector $\vm{x}$ is coded/compressed into a small dimensional vector $\vm{y}$ and 
the task will be to find the $\vm{x}$ from $\vm{y}$ with the full knowledge of $\vm{D}$. 
For this problem, the optimum solution is the sparsest vector satisfying 
\eqref{eq:Sparse_Representation_without_Noise}. Finding the sparsest vector is however 
NP-hard; thus, a variety of practical algorithms have been developed. Among the most 
prominent is the convex relaxation approach in which the objective is to find the 
minimum $l_{1}$-norm solution 
{to \eqref{eq:Sparse_Representation_without_Noise}}.
For the $l_{1}$-norm minimization, if $\vm{x}$ is $K$-sparse, 
{which indicates that the number of non-zero entries of $\vm{x}$ is at most $K$}, the minimum $K$ that satisfies \eqref{eq:Sparse_Representation_without_Noise} gives the limit up to which the signal can be compressed for a given dictionary $\vm{D}$. An interesting question then arises: How does the choice of the dictionary $\vm{D}$ affect the typical compression ratio that can be achieved using the $l_{1}$-recovery?

Recent results in the parallel problem of CS, where $\vm{D}$ acts as a sensing matrix, reveal that the typical conditions for perfect $l_{1}$-recovery are universal for all random sensing matrices that belong to the rotationally invariant matrix ensembles \cite{Kabashima-Wadayama-Tanaka-2009}. The standard setup, where the entries of the sensing matrix are independent standard Gaussian, is an example that belongs to this ensemble. It is also known that the conditions required for perfect recovery do not in general depend on the details of the marginal distribution related to the non-zero elements. On the other hand, we know that correlations in the sensing matrix can degrade the performance of $l_{1}$-recovery \cite{Takeda-Kabashima-isit2010}. This suggests intuitively that using a sample matrix of the rotationally invariant ensembles as $\vm{D}$ is preferred in the recovery problem when we expect to encounter a variety of dense signals $\vm{y}$. However, the set of matrix ensembles whose $l_1$-recovery performance are known is still limited, and further investigation is needed to assess whether the choice of $\vm{D}$ is indeed so straightforward.

The purpose of the present study is to fulfill this demand. Specifically, we examine the typical $l_1$-recovery performance of the matrices constructed by concatenating several randomly chosen orthonormal bases. Such construction has attracted considerable attention due to ease of implementation and theoretical elegance \cite{Donoho_Huo_2001}, \cite{Elad_Bruckstein_2002}, \cite{Donoho_Elad_2003} for designing sparsity inducing over-complete dictionaries for natural signals \cite{Rubinstein_Elad_Dictionary_ProcIEEE_2010}. For a practical engineering scheme, audio coding (music source coding) \cite{Ravelli_Richard_Daudet_2009_UnionMDCT} uses a dictionary formed by concatenating several modified discrete cosine transforms with different parameters.

By using the replica method in conjunction with the development of an integral formula for handling random orthogonal matrices, we show that the dictionary consisting of concatenated orthogonal matrices is also preferred in terms of the performance of $l_1$-recovery. More precisely, the matrices can result in better $l_1$-recovery performance than that of the rotationally invariant matrices when the density of non-zero entries of $\vm{x}$ is not uniform among the orthogonal matrix modules, while the performance is the same between the two types of matrices for the uniform densities. This surprising result further promotes the use of the concatenated orthogonal matrices in practical applications. 

This paper is organized as follows. In the next section, we explain 
the problem setting that we investigated. In Section 3, which is the main part of this paper, we discuss the development of a methodology for evaluating the recovery performance of the concatenated orthogonal matrices on the basis of the replica method and an integral formula concerning the random orthogonal matrices. In Section 4, we explain the significance of the methodology through application to two distinctive examples, the validity of which is also justified by extensive numerical experiments. The final section is devoted to a summary.

\section{Problem Setting}
We assume that $N$ is a multiple number of $M$; namely, $\nb=N/M =2,3,\ldots \in \mathbb{N}$. Suppose a situation in which an $M \times N$ dictionary matrix $\vm{D}$ is constructed by concatenating $T$ module matrices $\vm{O}_1, \vm{O}_2, \ldots, \vm{O}_\nb$, which are drawn uniformly and independently from the Haar measure on $M \times M$ orthogonal matrices, as
\begin{eqnarray}
\vm{D}=[\vm{O}_1\ \vm{O}_2\ \ldots \ \vm{O}_\nb]. 
\label{Lorth}
\end{eqnarray}
Using this, we compress a sparse vector $\vm{x}^0 \in \mathbb{R}^N$ to $\vm{y}\in \mathbb{R}^M$ following the manner of (\ref{eq:Sparse_Representation_without_Noise}). We denote $\vm{x}^0$ for the concatenation of $T$ sub-vectors of $M$ dimensions as
\begin{eqnarray}
\vm{x}^0=\left (
\begin{array}{c}
\vm{x}^0_1 \cr
\vm{x}^0_2 \cr
\vdots\cr
\vm{x}^0_\nb
\end{array}
\right ),
\label{vector_union}
\end{eqnarray}
yielding the expression
\begin{eqnarray}
\vm{y}=\vm{D}\vm{x}^0=\vm{O}_1 \vm{x}_1^0+\vm{O}_2 \vm{x}_2^0+ \ldots + \vm{O}_\nb \vm{x}_\nb^0.  
\label{compress}
\end{eqnarray}
With full knowledge of $\vm{O}_t$ $(t=1,2,\ldots,\nb)$ and $\vm{y}$, the $l_1$-recovery is performed by solving the constrained minimization problem 
\begin{eqnarray}
\mathop{\rm \min}_{\vm{x}_1,\vm{x}_2,\ldots,\vm{x}_\nb}
\left \{ \sum_{t=1}^\nb \|\vm{x}_t \|_{1} \right \} \ \mbox{subj. to} \ 
\vm{y}=\sum_{t=1}^\nb \vm{O}_t \vm{x}_t, 
\label{l1_recovery}
\end{eqnarray}
where 
{$\|\vm{x}\|_1=\sum_{i=1}^N |x_i|$ for $\forall\vm{x}=(x_i)\in \mathbb{R}^N$} and 
$\mathop{\rm \min}_{X}\left \{f(X) \right \}$ generally denotes the minimization of $f(X)$ with respect to $X$ and $\vm{x}_t \in \mathbb{R}^M$ $(t=1,2,\ldots,\nb)$. At the minimum condition, $\{\vm{x}_t\}$ constitutes the recovered vector $\hat{\vm{x}}$ in the manner of (\ref{vector_union}). 
 
For theoretically evaluating the $l_1$-recovery performance, we assume that the entries of $\vm{x}^0_t$, $x_{it}^0$ are distributed independently according to a block-dependent sparse distribution
\begin{eqnarray}
p_t(x)=(1-\rho_t)\delta(x)+\rho_t f_t(x), 
\label{sparse_dist}
\end{eqnarray}
where $0 \le \rho_t \le 1$ means the density of the non-zero entries of the $t$-th block of the same size $M$ in $\vm{x}^0$ and $f_t(x)$ is a distribution whose second moment about the origin is finite, which is assumed as unity for simplicity. Intuitively, as the compression rate $\alpha=M/N=\nb^{-1}$ decreases, the overall density $\rho=\nb^{-1}\sum_{t=1}^\nb \rho_t$ up to which (\ref{l1_recovery}) can successfully recover a typical sample of the original vector $\vm{x}^0$ becomes smaller. However, precise performance may depend on the profile of $\rho_t$. The above setting allows us to quantitatively examine how such block dependence of the non-zero density affects the critical relation between $\alpha$ and $\rho$ for typically successful $l_1$-recovery of $\vm{x}^0$.  

\section{Statistical mechanics approach}
\subsection{Statistical mechanical formulation}
Expressing the solution of (\ref{l1_recovery}) as
\begin{eqnarray}
\hat{\vm{x}}=\lim_{\beta \to \infty} \int d\vm{x}  P_\beta \left(\vm{x}|\vm{y},\{\vm{O}_t\} \right ) \vm{x}, 
\label{posterior_mean}
\end{eqnarray}
where
\begin{eqnarray}
&&P_\beta \left  (\vm{x}|\vm{y},\{\vm{O}_t\} \right )
=Z^{-1}(\beta;\{\vm{x}_t^0\}, \{\vm{O}_t\})\delta\left (\vm{y}-\sum_{t=1}^\nb \vm{O}_t \vm{x}_t \right ) 
\exp \left (-\beta \sum_{t=1}^\nb \|\vm{x}_t\|_1 \right ) \cr
&&=Z^{-1}(\beta;\{\vm{x}_t^0\}, \{\vm{O}_t\})
\delta\left (\sum_{t=1}^\nb \vm{O}_t (\vm{x}_t^0-\vm{x}_t) \right ) 
\exp \left (-\beta \sum_{t=1}^\nb \|\vm{x}_t\|_1 \right )
\label{posterior}
\end{eqnarray}
and $Z^{-1}(\beta;\{\vm{x}_t^0\}, \{\vm{O}_t\})\equiv \int \prod_{t=1}^\nb d\vm{x}_t \delta \left  (\sum_{t=1}^\nb \vm{O}_t (\vm{x}_t^0-\vm{x}_t) \right ) \exp \left (-\beta \sum_{t=1}^\nb \|\vm{x}_t\|_1 \right )$, constitutes the basis of our analysis. Equations (\ref{posterior_mean}) and (\ref{posterior}) mean that $\hat{\vm{x}}$ can be identified with the average of the state variable $\vm{x}$ for the Gibbs-Boltzmann distribution (\ref{posterior}) in the vanishing temperature limit $\beta \to \infty$. However, as (\ref{posterior}) depends on $\{\vm{x}_t^0\}$ and $\{\vm{O}_t\}$, further averaging with respect to the generation of these external random variables is necessary for evaluating the typical properties of the $l_1$-recovery. Evaluation of such ``double averages'' can be carried out systematically using the replica method \cite{replica}. 

\subsection{Integral formula for handling random orthogonal matrices}
In the replica method, we need to evaluate the average of $Z^n(\beta;\{\vm{x}_t^0\}, \{\vm{O}_t\})$ for $\forall{n} \in \mathbb{N}$ with respect to $\vm{O}_1, \vm{O}_2, \ldots,\vm{O}_\nb$ over the uniform distributions of $M \times M$ orthogonal matrices. However, this is rather laborious and is easy to yield notational confusions. For reducing such technical obstacles, we introduce a formula convenient for accomplishing this task before going into detailed manipulations. Similar formulae have been introduced for handling random eigenbases of symmetric matrices \cite{ItzksonZuber,MarinariParisiRitort,Dean,TakedaUdaKabashima,Kabashima-Wadayama-Tanaka-2009} and random left and right eigenbases of rectangular matrices \cite{Kabashima,ShinzatoKabashima}.

Let us assume that $M$-dimensional vectors $ \vm{u}_t=\vm{x}_t^0-\vm{x}_t$ $(t=1,2,\ldots,\nb)$ are characterized by their norms as
$v_1=M^{-1}|\vm{u}_1|^2, 
v_2=M^{-1}|\vm{u}_2|^2, 
\ldots, 
v_\nb=M^{-1}|\vm{u}_\nb|^2$, 
where $|\vm{u}|$ denotes the standard Euclidean norm of the vector $\vm{u}$. For these vectors, we define the function 
\begin{eqnarray}
F(\{v_t\})&=&\lim_{M \to \infty} 
\frac{1}{M}\ln \left (\left [ \delta \left (\sum_{t=1}^\nb \vm{O}_t \vm{u}_t \right ) \right ]_{\{\vm{O}_t\}} \right )\cr
&=&\lim_{M \to \infty} 
\frac{1}{M}\ln \left (\frac{\int \left (\prod_{t=1}^\nb {\cal D}\vm{O}_t \right )
\delta \left (\sum_{t=1}^\nb \vm{O}_t \vm{u}_t \right )}
{\int \left (\prod_{t=1}^\nb {\cal D}\vm{O}_t \right )}
\right ), 
\label{formula_def}
\end{eqnarray} 
where $\left [ f(X) \right ]_X$ generally denotes the average of $f(X)$ with respect to $X$, and ${\cal D}\vm{O}$ denotes the Haar measure of the $M \times M$ orthogonal matrices. 
Our claim is that by explicitly using $v_1, v_2, \ldots, v_\nb$, \eqref{formula_def} can be expressed as
\begin{eqnarray}
&&F(\{v_t\}) = \mathop{\rm extr}_{\Lambda_1,\Lambda_2,\ldots,\Lambda_\nb}
\left \{
-\frac{1}{2}\ln \left (\sum_{t=1}^\nb \Lambda_t^{-1} \right )
-\frac{1}{2}\sum_{t=1}^{\nb}  \ln (\Lambda_t) 
 + \frac{1}{2}\sum_{t=1}^{\nb}  \Lambda_t v_t \right \} \cr
&& \hspace*{6.5cm}-\frac{1}{2} \sum_{t=1}^{\nb} \ln (v_t)-\frac{\nb}{2}, 
\label{formula_concrete}
\end{eqnarray}
where $\mathop{\rm extr}_{X} \left \{ f(X) \right \}$ generally denotes the extremization of function $f(X)$ with respect to $X$. Expression (\ref{formula_concrete}) is derived from the fact that for fixed $\vm{u}_t$, $\vm{O}_t \vm{u}_t$ moves uniformly on the surface of the $M$-dimensional hypersphere of radius $\sqrt{M v_t}$ when $\vm{O}_t$ varies according to the uniform distribution of the orthogonal matrices; therefore, ${\cal D}\vm{O}_t$ in (\ref{formula_def}) can be replaced with a spherical measure of $M$-dimensional vector $ d \vm{u}_t \delta (|\vm{u}_t|^2-Mv_t)$. For details, see \ref{appendix1}. 


Function $F(\{v_t\})$ physically represents a characteristic exponent of the probability that $M$-dimensional vectors $\vm{u}_t$ $(t=1,2,\ldots,\nb)$ form a closed loop satisfying $\sum_{t=1}^\nb \vm{u}_t=\vm{0}$ when they are independently and isotropically sampled under the norm constraints of $|\vm{u}_t|^2=Mv_t$. For small $T$, the loop condition strongly restricts the region of $\{v_t\}$ to which $F(\{v_t\})$ is well defined. In concrete terms, $F(\{v_t\})$ diverges to minus infinity unless an equality  
\begin{eqnarray}
v_1=v_2 
\label{T2}
\end{eqnarray}
and the triangular inequalities
\begin{eqnarray}
v_1 < v_2+v_3, \ v_2 < v_1+v_3, \ v_3 < v_1+v_2
\label{T3}
\end{eqnarray}
are satisfied for  $\nb=2$ and $3$, respectively. In particular, the constraint of (\ref{T2}) requires us to deal with the system of $\nb=2$ in a manner different from that of $\nb \ge 3$ except for the case of $\rho_1=\rho_2$, as discussed in the analysis 
of section 3.4.

\subsection{Replica method}
Now, we are ready to apply the replica method for analyzing the typical property of the $l_1$-recovery (\ref{posterior_mean}). For this, we evaluate the $n$-th moment of the partition function using the identity
\begin{eqnarray}
&&Z^n(\beta;\{\vm{x}_t^0\}, \{\vm{O}_t\})
=\int \left (\prod_{a=1}^n \prod_{t=1}^\nb d\vm{x}_t^a \right ) 
\left (\prod_{a=1}^n 
\delta\left (\sum_{t=1}^\nb \vm{O}_t (\vm{x}_t^0-\vm{x}_t^a) \right ) \right ) \cr
&& \hspace*{4cm}
\times
\exp \left (-\beta \sum_{a=1}^n \sum_{t=1}^\nb \|\vm{x}_t^a\|_1 \right ), 
\label{power_partition}
\end{eqnarray}
which is valid for only $n \in \mathbb{N}$. In the large system limit $M \to \infty$, 
the rescaled logarithm of the moment, $M^{-1} \ln \left [Z^n(\beta;\{\vm{x}_t^0\}, 
\{\vm{O}_t\}) \right ]_{\{\vm{x}_t^0\}, \{\vm{O}_t\}}$, can be accurately evaluated for 
{all $n \in \mathbb{N}$} by using the saddle point method with 
respect to macroscopic variables $Q_t^{a}=M^{-1} |\vm{x}_t^a|^2$, $q_t^{ab}=M^{-1} 
\vm{x}_t^a \cdot \vm{x}_t^b$, and $m_t^a=M^{-1}\vm{x}_t^0 \cdot \vm{x}_t^a$, where 
$a,b=1,2,\ldots,n$ and $t=1,2,\ldots,\nb$. Intrinsic permutation symmetry concerning the 
replica indices $a=1,2,\ldots, n$ in (\ref{power_partition}) guarantees that there 
exists a saddle point of the form $Q_t^a=Q_t$, $q_t^{ab}=q_t$, and $m_t^a=m_t$, which is 
often termed the replica symmetric (RS) solution. As a simple and plausible candidate, 
we adopt this solution as the relevant saddle point for describing the typical property 
of the $l_1$-recovery, the validity of which will be checked 
in section 3.5. 
The detailed computation is carried out as follows. 

\subsubsection{Energetic part}
{Let us consider averaging (\ref{power_partition}) with
respect to $\{\vm{O}_t\}$ and define for each fixed set of $\{\vm{x}_t^a\}$:}
\begin{eqnarray}
{\cal I}(\{\vm{u}_t^a\}) 
= \left [ \prod_{a=1}^n 
\delta\left (\sum_{t=1}^\nb \vm{O}_t \vm{u}_t^a \right ) \right ]_{\{\vm{O}_t\}}, 
\label{average_O}
\end{eqnarray}
where $\vm{u}_t^a=\vm{x}_t^0 - \vm{x}_t^a$. When $\{\vm{x}_t^a\}$ is placed in the configuration of the RS solution, the expression 
\begin{eqnarray}
&&\left [ \vm{u}_t^1 \ \vm{u}_t^2 \ \ldots \ \vm{u}_t^n \right ]^{\rm T} \times 
\left [ \vm{u}_t^1 \ \vm{u}_t^2 \ \ldots \ \vm{u}_t^n \right ]
=\left (
\begin{array}{cccc}
MR_t & Mr_t &  \cdots & Mr_t \cr
Mr_t & MR_t &  \cdots & Mr_t \cr
\vdots & \vdots & \ddots & \vdots \cr
Mr_t & Mr_t & \cdots & M R_t 
\end{array}
\right ) \cr
&&= \vm{E} \times 
\left ( 
\begin{array}{cccc}
M(R_t-r_t+nr_t) & 0 & \cdots & 0 \cr
0 & M(R_t-r_t) & \cdots & 0 \cr
\vdots & \vdots & \ddots & \vdots \cr
0 & 0 & \cdots & M(R_t-r_t) 
\end{array}
\right ) \times \vm{E}^{\rm T}
\end{eqnarray}
holds for each $t$, where $\rm{T}$ stands for matrix transpose, $R_t=Q_t-2m_t+\rho_t$, and $r_t=q_t-2m_t+\rho_t$. $\vm{E}=[\vm{e}_1 \ \vm{e}_2 \ \ldots \ \vm{e}_n]$ denotes an $n \times n$ orthogonal matrix composed of the vector $\vm{e}_1=(n^{-1/2},n^{-1/2},\ldots, n^{-1/2})^{\rm T}$ and an orthonormal set of $n-1$ vectors $\vm{e}_2, \vm{e}_3, \ldots, \vm{e}_n$ that are orthogonal to $\vm{e}_1$. This indicates that $\left [ \vm{u}_t^1 \ \vm{u}_t^2 \ \ldots \ \vm{u}_t^n \right ]$ may be expressed as
\begin{eqnarray}
\left [ \vm{u}_t^1 \ \vm{u}_t^2 \ \ldots \ \vm{u}_t^n \right ]=
\left [\tilde{\vm{u}}_t^1\  \tilde{\vm{u}}_t^2 \ \ldots \ \tilde{\vm{u}}_t^n \right ] \times \vm{E}^{\rm T}, 
\label{cordinate_conversion}
\end{eqnarray}
by using a set of $n$ orthogonal vectors $\{\tilde{\vm{u}}_t^a\}$, whose norms are given as $|\tilde{\vm{u}}_t^1|^2=M(R_t-r_t+nr_t)=M(Q_t-q_t+n(q_t-2m_t+\rho_t))$ and $|\tilde{\vm{u}}_t^a|^2=M(R_t-r_t)=M(Q_t-q_t)$ for $a=2,3,\ldots, n$, along with an $n\times n$ orthogonal matrix $\vm{E}$ that does not depend on $t$. This guarantees the equality ${\cal I}(\{\vm{u}_t^a\}) ={\cal I}(\{\tilde{\vm{u}}_t^a\})$. Furthermore, condition $n \ll M$ and the orthogonality of $\{\tilde{\vm{u}}_t^a\}$ among the replica indexes $a=1,2,\ldots, n$ allows us to evaluate the average concerning $\{\vm{O}_t\}$ independently for each index $a$ when computing ${\cal I}(\{\tilde{\vm{u}}_t^a\})$. This, in conjunction with (\ref{formula_def}), provides each set $\{\vm{x}_t^a\}$ of the RS configuration with an expression of (\ref{average_O}) as 
\begin{eqnarray}
\frac{1}{M} \ln {\cal I}(\{\vm{u}_t^a\}) 
=F(\{Q_t-q_t+n(q_t-2m_t+\rho_t)\})+(n-1)F(\{Q_t-q_t\}).
\label{energy}
\end{eqnarray}
The right hand side of (\ref{energy}) is likely to hold for $n \in \mathbb{R}$ as well, although (\ref{average_O}) is defined originally for only $n \in \mathbb{N}$. 

\subsubsection{Entropic part}
On the other hand, inserting identities $1=\int MdQ_t \delta (|\vm{x}_t^a|^2-MQ_t)$, $1=\int Mdq_t \delta (\vm{x}_t^a \cdot \vm{x}_t^b -Mq_t)$ and $1=\int Mdm_t \delta (\vm{x}_t^0 \cdot \vm{x}_t^a -Mm_t)$ $(a,b=1,2,\ldots,n; t=1,2,\ldots, \nb)$ into (\ref{power_partition}) and taking an average concerning $\{\vm{x}_t^0\}$, in conjunction with integration with respect to dynamical variables $\{\vm{x}_t^a\}$, result in the expression 
\begin{eqnarray}
&&{\cal V}(\{Q_t,q_t,m_t\})=\int \left (\prod_{t=1}^\nb \prod_{a=1}^n 
d\vm{x}_t^a \exp \left (-\beta \|\vm{x}_t^a\|_1\right )\right ) \times 
\left (\prod_{t=1}^\nb \prod_{a=1}^n  \delta (|\vm{x}_t^a|^2-MQ_t)  \right )\cr
&& \hspace*{2cm}
\times \left (\prod_{t=1}^\nb \prod_{a>b}  \delta (\vm{x}_t^a \cdot \vm{x}_t^b -Mq_t) \right )
\times \left (\prod_{t=1}^\nb \left [\prod_{a=1}^n \delta (\vm{x}_t^0 \cdot \vm{x}_t^a -Mm_t) \right ]_{\vm{x}_t^0}\right ) \cr
&&=
\int d \hat{\vm{Q}} 
\exp \left (M (A(\{Q_t,q_t,m_t\}, \{\hat{Q}_t^a,\hat{q}_t^{ab},\hat{m}_t^a\})
+B(\{\hat{Q}_t^a,\hat{q}_t^{ab},\hat{m}_t^a\}) \right ),  
\label{volume}
\end{eqnarray} 
for a fixed set of $\{Q_t,q_t,m_t\}$. The conjugate variable $\hat{Q}_t^a$ is introduced for expressing a delta function as $ \delta (|\vm{x}_t^a|^2-MQ_t)=(4\pi \sqrt{-1})^{-1} \int_{-\sqrt{-1}\infty}^{+\sqrt{-1}\infty} d \hat{Q}_t^a \exp \left (-(\hat{Q}_t^a/2) \left (|\vm{x}_t^a|^2-MQ_t \right ) \right )$, and similarly for $\hat{q}_t^{ab}$ and $\hat{m}_t^a$. Notation $d\hat{\vm{Q}}$ stands for an integral measure $\prod_{t=1}\left (\prod_{a=1}d\hat{Q}_t^a \prod_{a>b}d\hat{q}_t^{ab} \prod_{a=1}d\hat{m}_t^a \right )$, and the functions on the right hand side are defined as 
\begin{eqnarray}
A(\{Q_t,q_t,m_t\}, \{\hat{Q}_t^a,\hat{q}_t^{ab},\hat{m}_t^a\})
=\sum_{t=1}^\nb
\left (
\sum_{a=1}^n \frac{\hat{Q}_t^aQ_t}{2}
-\sum_{a>b} \hat{q}_t^{ab}q_t
-\sum_{a=1}^n \hat{m}_t^a m_t
\right )
\end{eqnarray}
and 
\begin{eqnarray}
&&B(\{\hat{Q}_t^a,\hat{q}_t^{ab},\hat{m}_t^a\})
=\sum_{t=1}^\nb 
\ln 
\left (
\left [
\int \left (\prod_{a=1}^n dx_t^{a} \right )
\exp \left (-\sum_{a=1}^n \frac{\hat{Q}_t^a}{2} (x_t^a)^2 
+ \sum_{a>b} \hat{q}_t^{ab} x_t^a x_t^b \right . \right . \right . \cr
&& \hspace*{7cm}\left . \left . \left . 
+ \sum_{a=1}^n \hat{m}_t^a x_t^0 x_t^a-\sum_{a=1}^n 
\beta |x_t^a| 
\right )
\right ]_{x_t^0} 
\right ), 
\end{eqnarray}
where the average for $x_t^0$ is taken according to (\ref{sparse_dist}). The expression of (\ref{volume}) indicates that its rescaled logarithm is accurately evaluated using the saddle point method with respect to the conjugate variables in the large system limit $M \to \infty$. In addition, the replica symmetry guarantees that the relevant saddle point is of the RS form as $\hat{Q}_t^a=\hat{Q}_t$, $\hat{q}_t^{ab}=\hat{q}_t$, and $\hat{m}_t^a=\hat{m}_t$. As a consequence, the evaluation yields
\begin{eqnarray}
&&\frac{1}{M}\ln {\cal V}(\{Q_t,q_t,m_t\})
=\sum_{t=1}^\nb
\mathop{\rm extr}_{\hat{Q}_t,\hat{q}_t,\hat{m}_t}
\left \{
\frac{n\hat{Q}_t Q_t}{2}-\frac{n(n-1) \hat{q}_t q_t}{2} -n \hat{m}_t m_t \right . \cr
&& \left . + \ln \left (
\!\left [\!
\int Dz\! 
\left (
\!\int dx 
\exp \left (\!-\frac{(\hat{Q}_t\!+\!\hat{q}_t)x^2}{2}\!+\!(\!\sqrt{\hat{q}_t}z\!+\!\hat{m}_t x_t^0)x\! -\!\beta |x| \! \right )
\right )^n
\right ]_{x_t^0} \!\right )
\!\right \}, 
\label{entropy}
\end{eqnarray}
where $Dz=dz\exp (-z^2/2)/\sqrt{2\pi}$ denotes the Gaussian measure. This is also likely to hold for $n \in \mathbb{R}$, although (\ref{volume}) is originally defined for only $n \in \mathbb{N}$.

\subsubsection{Free energy and saddle point equations}
The replica method uses the identity $-(\beta M)^{-1}\left [ \ln Z(\beta;  \{\vm{x}_t^0\}, \{\vm{O}_t\}) \right ]_{\{\vm{x}_t^0\}, \{\vm{O}_t\}} =-\lim_{n\to 0} (\partial /\partial n)(\beta M)^{-1} \ln \left (\left [ Z^n(\beta;\{\vm{x}_t^0\}, \{\vm{O}_t\}\right . \right . $ $\left . \left . ) \right ]_{\{\vm{x}_t^0\}, \{\vm{O}_t\}} \right )$ for evaluating the typical free energy density. The above argument indicates that $M^{-1} \ln \left (\left [ Z^n(\beta;\{\vm{x}_t^0\}, \{\vm{O}_t\}) \right ]_{\{\vm{x}_t^0\}, \{\vm{O}_t\}} \right )$ can be computed by extremizing the sum of (\ref{energy}) and (\ref{entropy}) with respect to $\{Q_t,q_t,m_t\}$. Furthermore, the obtained expression is likely to hold for $n \in \mathbb{R}$, although the calculations are based on (\ref{power_partition}) that is valid for only $n \in \mathbb{N}$. 
We, therefore, take the limit of $n\to 0 $ utilizing the expressions of (\ref{energy}) and (\ref{entropy}) for $n \in [0,1]$ as well.  
In particular, in the limit of $\beta \to \infty$, which is relevant in the current problem, the expression of the free energy of the vanishing temperature is expressed as
\begin{eqnarray}
&&-\lim_{\beta \to \infty}
\frac{1}{\beta M} \left [ \ln Z(\beta;\{\vm{x}_t^0\}, \{\vm{O}_t\}) \right ]_{\{\vm{x}_t^0\}, \{\vm{O}_t\}} \cr
&&= -\mathop{\rm extr}_{}\left \{\sum_{t=1}^\nb 
\left (\frac{\partial F(\{\chi_k\})}{\partial \chi_t} (Q_t-2 m_t+\rho_t) 
+\frac{\hat{Q}_tQ_t}{2}-\frac{\hat{\chi}_t\chi_t}{2}+\hat{m}_tm_t \right . \right .\cr
&&\left . \left . \hspace*{6cm} -\int Dz \left [ \phi \left (\sqrt{\hat{\chi}_t} z+ 
{\hat{m}_t} x^0; \hat{Q}_t \right ) \right ]_{x^0}
\right ) \right \}, 
\label{free_energy}
\end{eqnarray}
where 
\begin{eqnarray}
\phi\left (\sqrt{\hat{\chi}_t} z+ 
{\hat{m}_t} x^0; \hat{Q}_t \right )
= \mathop{\rm min}_{x} \left \{
\frac{\hat{Q}_t}{2}x^2-\left (\sqrt{\hat{\chi}_t} z+ \hat{m}_t x^0 \right )x+|x|
\right \}, 
\label{phi}
\end{eqnarray}
and rescaled variables are introduced as $\beta (Q_t-q_t) \to \chi_t$, $\hat{q}_t/\beta^2 \to \hat{\chi}_t$, $(\hat{Q}_t+\hat{q}_t) /\beta \to \hat{Q}_t$, and $\hat{m}_t/\beta \to \hat{m}_t$ to properly describe the relevant solution in the limit of $\beta \to \infty$. 
Extremization is to be performed with respect to $\{Q_t,\chi_t,m_t, \hat{Q}_t, \hat{\chi}_t, \hat{m}_t\}$. 

Similar to earlier studies, at the extremum characterized by a set of the saddle point equations
\begin{eqnarray}
&&\hat{Q}_t=-2 \frac{\partial F(\{\chi_k\})}{\partial \chi_t} =\frac{R_t}{\chi_t}, \label{SPh1}\\
&&\hat{\chi}_t=2\sum_{s=1}^\nb \frac{\partial^2 F( \{\chi_k\})}{\partial \chi_t\partial \chi_s}
(Q_s-2m_s+\rho_s), \label{SPh2} \\
&&\hat{m}_t=-2 \frac{\partial F(\{\chi_k\})}{\partial \chi_t} =\frac{R_t}{\chi_t}, \label{SPh3}
\end{eqnarray}
\begin{eqnarray}
&&Q_t=\int Dz \left [ X^2(\sqrt{\hat{\chi}_t}z+\hat{m}_t x_t^0;\hat{Q}_t) \right ]_{x_t^0}, \label{SP1}\\
&&\chi_t=\int Dz \left [ \frac{\partial X^2(\sqrt{\hat{\chi}_t}z+\hat{m}_t x_t^0;\hat{Q}_t)}{\partial (\sqrt{\hat{\chi}_t}z) }
\right ]_{x_t^0}, \label{SP2}\\
&& m_t=\int Dz \left [x_t^0X(\sqrt{\hat{\chi}_t}z+\hat{m}_t x_t^0;\hat{Q}_t) \right ]_{x_t^0}, \label{SP3}
\end{eqnarray}
$Q_t$ and $m_t$ $(t=1,2,\ldots,\nb)$ physically denote the macroscopic averages of the recovered vector $\hat{\vm{x}}_t$ as $M^{-1} \left [ |\hat{\vm{x}}_t|^2 \right ]_{\{\vm{x}_k^0\},\{\vm{O}_k\}}$ and $M^{-1} \left [\vm{x}_t^0 \cdot \hat{\vm{x}}_t \right ]_{\{\vm{x}_k^0\},\{\vm{O}_k\}}$, respectively. Here, $R_t=\Lambda_t^{-1}/\left (\sum_{t=1}^\nb \Lambda_t^{-1} \right )$ is provided by the extremum solution of (\ref{formula_concrete}) for $\{v_t\}=\{\chi_t\}$, and 
\begin{eqnarray}
X(h;\hat{Q})=-\partial \phi(h;\hat{Q})/\partial h 
=
\left \{
\begin{array}{ll}
(h-1)/{\hat{Q}}, & h> 1,\cr
0, & |h| \le 1, \cr
(h+1)/{\hat{Q}}, & h < -1.  
\end{array}
\right . 
\label{recover_func}
\end{eqnarray}
For $T=2$, a Lagrange multiplier should be exceptionally introduced in (\ref{SPh2}) for enforcing $\chi_1=\chi_2$. 

\subsection{Critical condition for $l_1$-recovery}
The success of the $l_1$-recovery is characterized by the condition in which $Q_t=m_t=\rho_t$ is satisfied at the extremum for $\forall{t} =1,2,\ldots,\nb$. Therefore, one can evaluate the critical relation between $\alpha=1/\nb$ and $\rho=\nb^{-1}\sum_{t=1}^T \rho_t$ by examining the thermodynamic stability of the success solution $Q_t=m_t=\rho_t$ $(t=1,2,\ldots,\nb)$ for the saddle point equations (\ref{SPh1})--(\ref{SP3}).

We assume $T\ge 3$ for a while, since an exceptional treatment is required for $T=2$. For obtaining the success solution, it is necessary that $\chi_t = 0$ and $\hat{Q}_t=\hat{m}_t = +\infty$ hold. Expanding (\ref{SP1})--(\ref{SP3}) under the assumption of $\hat{Q}_t=\hat{m}_t \gg1$ yields
\begin{eqnarray}
Q_t\simeq \rho_t-\frac{4 \rho_t}{\sqrt{2 \pi}\hat{Q}_t}+\frac{1}{\hat{Q}_t^2}
\left ( 2(1-\rho_t) \G(\hat{\chi}_t)+\rho_t(\hat{\chi}_t+1) \right )+O(\hat{Q}_t^{-3}), \label{Qasympt}\\
m_t \simeq \rho_t -\frac{2 \rho_t}{\sqrt{2 \pi}\hat{Q}_t}+O(\hat{Q}_t^{-3}), \label{masympt}\\
\chi_t \simeq \frac{2 (1-\rho_t)\Q(\hat{\chi}_t^{-1/2})+\rho_t}{\hat{Q}_t}, \label{chiasympt}
\end{eqnarray}
where $\Q(x)=\int_x^{\infty} Dz$, 
\begin{eqnarray}
\G(x)=(x+1)\Q(x^{-1/2})-x^{1/2}\frac{{ e}^{-1/(2x)}}{\sqrt{2\pi}}, 
\label{Gfunc}
\end{eqnarray}
and we used $[(x_t^0)^2]_{x_t^0}=\rho_t$, which is derived from the assumption (\ref{sparse_dist}). These result in the expression
\begin{eqnarray}
Q_t-2m_t+\rho_t \simeq \frac{2(1-\rho_t) \G(\hat{\chi}_t)+\rho_t(\hat{\chi}_t+1) }{\hat{Q}_t^2} +
O(\hat{Q}_t^{-3}).
\label{MSE}
\end{eqnarray}

In addition, (\ref{formula_concrete}) indicates that 
\begin{eqnarray}
\frac{\partial^2F(\{\chi_k\})}{\partial \chi_t \partial \chi_s}=\frac{1}{2}
\frac{\partial \Lambda_s}{\partial \chi_t}+\frac{1}{2 \chi_t^2}\delta_{ts} 
=\frac{1}{2}\left (\frac{\partial \chi_m}{\partial \Lambda_n} \right )^{-1}_{ts}
+\frac{1}{2 \chi_t^2}\delta_{ts} 
\label{hessian}
\end{eqnarray}
holds, where $\{\Lambda_t\}$ is the solution of $T$ coupled equations 
\begin{eqnarray}
\chi_t=\frac{1}{\Lambda_t}\left (1-\frac{\Lambda_t^{-1}}{\sum_{t=1}^\nb \Lambda_t^{-1} } \right )
=\frac{1}{\Lambda_t}(1-R_t).
\label{chi_lambda}
\end{eqnarray}
Differentiating this yields
\begin{eqnarray}
\frac{\partial \chi_m}{\partial \Lambda_n}
=-\frac{(1-2R_m)}{\Lambda_m^2} \delta_{mn}-\frac{R_mR_n}{\Lambda_m\Lambda_n}
\equiv -(J_{mn}+K_{mn}), 
\end{eqnarray} 
where we denote $\vm{J}=(J_{mn})=\left ((1-2R_m)\Lambda_m^{-2} \delta_{mn} \right )$ and $\vm{K}=(K_{mn})=\left ((R_mR_n)(\Lambda_m\Lambda_n)^{-1}\right).$ This expression makes it possible to evaluate $(\partial \Lambda_m/\partial \chi_n)=(\partial \chi_m/\partial \Lambda_n)^{-1}$ as
\begin{eqnarray}
\left (\frac{\partial \Lambda_m}{\partial \chi_n} \right )
&=&\left (\frac{\partial \chi_m}{\partial \Lambda_n} \right )^{-1}=
-(\vm{J}(\vm{I}+\vm{J}^{-1}\vm{K}) )^{-1}=-(\vm{I}+\vm{J}^{-1}\vm{K} )^{-1}\vm{J}^{-1}\cr
&=& -\sum_{p=0}^\infty (-1)^p (\vm{J}^{-1}\vm{K})^p \vm{J}^{-1} = -\vm{T}(\vm{I}+\vm{M})^{-1}\vm{T}, 
\label{hessian_inverse}
\end{eqnarray}
where $\vm{I}=(\delta_{mn})$, $\vm{T}=(\Lambda_m(1-2 R_m)^{-1/2} \delta_{mn})$ and $\vm{M}=(R_m R_n (1-2R_m)^{-1/2}(1-2R_n)^{-1/2})$. Inserting (\ref{MSE}), (\ref{hessian}), and (\ref{hessian_inverse}) into (\ref{SPh2}) yields a set of equations to determine $\{\hat{\chi}_t\}$ for a given set of non-zero densities $\{\rho_t\}$
\begin{eqnarray}
&&\hat{\chi}_t=\frac{2 (1-\rho_t)\G(\hat{\chi}_t)+\rho_t(\hat{\chi}_t+1)}{R_t^2}\cr
&&\hspace*{1cm}-\sum_{s=1}^\nb
(\vm{S}(\vm{I}+\vm{M})^{-1}\vm{S})_{ts}(1-R_s)^2 
\left (2(1-\rho_s)\G(\hat{\chi}_s)+\rho_s(\hat{\chi}_s+1) \right ), 
\label{critical1}
\end{eqnarray}
where we set $\vm{S}=(\delta_{mn}/(R_m\sqrt{1-2 R_m}))$ and used the relation $\vm{T}/\hat{Q}_t=(1-R_t)\vm{S}$, which is obtained from (\ref{SPh1}) and (\ref{chi_lambda}).

Equations (\ref{SPh2}) and (\ref{chiasympt}) indicate that the critical condition for making $\chi_t=0$ stable is expressed as
\begin{eqnarray}
R_t=2(1-\rho_t)\Q(\hat{\chi}_t^{-1/2})+\rho_t. 
\label{critical2}
\end{eqnarray}
Furthermore, the condition 
\begin{eqnarray}
\sum_{t=1}^\nb R_t =1
\label{critical3}
\end{eqnarray}
must hold by the definition of $R_t$. 

For characterizing the critical condition for the success of the $l_1$-recovery, let us suppose that the set of non-zero densities $\{\rho_t\}$ is provided as a function of a single parameter $\mu$ as $\{\rho_t(\mu)\}$. For $T \ge 3$, the critical situation is specified by an appropriate set of $2\nb+1$ variables of $\mu$, $\{\hat{\chi}_t\}$, and $\{R_t\}$. These are provided by $2\nb+1$ conditions of (\ref{critical1})--(\ref{critical3}). 

On the other hand, the critical condition for $T=2$ is provided differently from (\ref{critical1})--(\ref{critical3}) because the constraint of $\chi_1=\chi_2$ for keeping $F(\{\chi_1,\chi_2\})$ well defined requires $R_1=R_2=1/2$ for any pair of $\rho_1$ and $\rho_2$. Explicitly, the condition is provided by the following four coupled equations:
\begin{eqnarray}
\hat{\chi}_1=4(1-\rho_1(\mu))\G(\hat{\chi}_1)+4\rho_1(\mu)(\hat{\chi}_1+1)+\eta, 
\label{T2critical1} \\
 \hat{\chi}_2=4(1-\rho_2(\mu))\G(\hat{\chi}_2)+4\rho_2(\mu)(\hat{\chi}_2+1)-\eta, 
\label{T2critical2} \\
2(1-\rho_1(\mu))\Q(\hat{\chi}_1^{-1/2})+2 \rho_1(\mu)=\frac{1}{2}, 
\label{T2critical3} \\ 
2(1-\rho_2(\mu))\Q(\hat{\chi}_2^{-1/2})+2 \rho_2(\mu)=\frac{1}{2}, 
\label{T2critical4}
\end{eqnarray}
where $\eta$ is a Lagrange parameter for enforcing $\chi_1=\chi_2$. These determine four variables of $\hat{\chi}_1,\hat{\chi}_2, \eta, \mu$ at the critical condition.  

Equations (\ref{critical1})--(\ref{critical3}) for $T\ge 3$ and (\ref{T2critical1})--(\ref{T2critical4}) for $T=2$ constitute the main result of this paper. 
{For $T=2$, an equivalent result can also be obtained in a slightly different manner \cite{Vehkapera_ITW2012}}.

\subsection{Validity of RS evaluation}
The above calculation can be generalized to arbitrary levels of replica symmetry breaking (RSB). For example, under one-step RSB (1RSB) ansatz, where $n$ replicas of each block $t$ are classified into $n/x$ groups of an identical size $x$ and their overlaps are assumed to be $\vm{x}_t^a\cdot\vm{x}_t^b/M =q_{1t}$ if $a$ and $b$ belong to the same group and $q_{0t}(\le q_{1t})$ otherwise, (\ref{energy}) and (\ref{entropy}) are modified as
\begin{eqnarray}
&&\frac{1}{M} \ln {\cal I}(\{\vm{u}_t^a\}) 
=F(\{Q_t-q_{1t}+x(q_{1t}-q_{0t})+n(q_{0t}-2m_t+\rho_t)\})\cr
&&+\left (\frac{n}{x}-1 \right )F(\{Q_t-q_{1t}+x(q_{1t}-q_{0t})\}) +\left (n-\frac{n}{x} \right )F(\{Q_t-q_{1t}\})
\label{energy1RSB}
\end{eqnarray}
and 
\begin{eqnarray}
&&\frac{1}{M}\ln {\cal V}(\{Q_t,q_{1t},q_{0t},m_t\})
=\sum_{t=1}^\nb
\mathop{\rm extr}_{\hat{Q}_t,\hat{q}_{1t},\hat{q}_{0t},\hat{m}_t}
\left \{
\frac{n\hat{Q}_t Q_t}{2}\!  -\!\frac{n(x-1)(\hat{q}_{1t}q_{1t}\!-\!\hat{q}_{0t} q_{0t} )}{2} \right . \cr
&&\left . -\!\frac{n(n-1) \hat{q}_{0t} q_{0t}}{2} \!-\!n \hat{m}_t m_t  + \ln \left (
\!\left [\!
\int \!\! Dz_0\!\! 
\left ( \! \int \!\!Dz_1\!\! 
\left (\!\int \! dx \exp (\Phi )
 \!\right )^x
\! \right )^{n/x}
\! \right )\!\right ]_{x_t^0}
\!\right \}, 
\label{entropy1RSB}
\end{eqnarray}
respectively, where $\Phi =-(\hat{Q}_t+\hat{q}_{1t})x^2/2+(\sqrt{\hat{q}_{1t}-\hat{q}_{0t}}z_1+\sqrt{\hat{q}_{0t}}z_0 +\hat{m}_tx_t^0)x-\beta|x|$. 

In the 1RSB framework, the RS solution is regarded as a special solution of $\Delta_t=q_{1t}-q_{0t}=0$ and $\hat{\Delta}_t=\hat{q}_{1t}-\hat{q}_{0t}=0$  ($t=1,2,\ldots,\nb$). In the limit of $n \to 0$ and $\beta \to \infty$ keeping $\chi=\beta(Q_t-q_{1t}) \sim O(1)$, the critical condition that a solution of $\Delta_t>0$ and $\hat{\Delta}_t >0$ bifurcates from the RS solution, which corresponds to the de Almeida-Thouless (AT) condition \cite{AT} of the current system, is expressed as
\begin{eqnarray}
\det \left (\vm{I}-\vm{H} \hat{\vm{H}} \right )=0, 
\label{AT}
\end{eqnarray}
where $\vm{H}=(2 (\partial^2/\partial \chi_t \partial \chi_s) F(\{\chi_k\}))$ and $\hat{\vm{H}}=( \delta_{ts} \int Dz [(\partial X
(\sqrt{\hat{\chi_t}}z+\hat{m}_t x_t^0;\hat{Q}_t) /\partial(\sqrt{\hat{\chi_t}}z))^2 ]_{x_t^0} )$. 

For $T \ge 3$, 
\begin{eqnarray}
\vm{H}\hat{\vm{H}}=\left (\frac{1}{R_t}\delta_{ts} 
-\frac{(\vm{I}+\vm{M})^{-1}_{ts}(1-R_s)^2}{R_t\sqrt{1-2R_t}\sqrt{1-2R_s}} \right ) 
\label{HhatH}
\end{eqnarray}
holds for the success solution at the critical condition of the $l_1$-recovery. Equation (\ref{HhatH}) yields an eigenvalue of unity whose eigenvector is given as $\vm{v}_1 \propto ((1-R_t)^{-1})$, which makes (\ref{AT}) hold. Similarly, (\ref{AT}) is also satisfied at the critical condition of the $l_1$-recovery of $T=2$. These validate our RS evaluation in terms of the local stability analysis, although further justification with other schemes, such as comparison with numerical experiments, is necessary for examining possibilities that the RS solution becomes thermodynamically irrelevant due to discontinuous phase transitions.

\section{Case studies}
Let us examine the significance of the developed methodology by applying it to two representative examples.  

\subsection{Uniform and localized densities}
We consider the uniform density case of $\rho_t=\rho$ ($t=1,2,\ldots,\nb$) as the first example, where the uniformity allows us to solve (\ref{SPh1})--(\ref{SP3}) setting all variables to be independent of $t$ as $\chi_t=\chi$. In particular, setting $\Lambda_t=\Lambda$ simplifies the expressions of (\ref{SPh1})--(\ref{SPh3}) providing $R_t=1/T$ and $2 (\partial^2/\partial \chi_t\partial \chi_s  ) F( \{\chi_k\})=\chi^{-2}$. This makes it unnecessary to deal with the saddle point problems of $T=2$ in an exceptional manner. As a consequence, the critical condition of the $l_1$-recovery is expressed compactly by using a pair of equations as 
\begin{eqnarray}
\hat{\chi}&=&T\left (2 (1-\rho)\G(\hat{\chi})+\rho (\hat{\chi}+1) \right ), 
\label{uniform1}\\
T^{-1}&=&2(1-\rho) \Q(\hat{\chi}^{-1/2})+\rho, 
\label{uniform2}
\end{eqnarray}
for both $T\ge3$ and $T=2$. By setting $\alpha=1/T$, these provide a critical condition identical to that obtained for the rotationally invariance matrix ensembles in earlier studies \cite{Kabashima-Wadayama-Tanaka-2009,Donoho-2006-geometry,DonohoTanner}. This indicates that for vectors of the uniform non-zero density, the $l_1$-recovery performance of the concatenated orthogonal matrices is identical to that of the standard setup provided by the matrix of independent standard Gaussian entries.   

\begin{figure}
\centering
\includegraphics[width=0.5\columnwidth,angle=270]{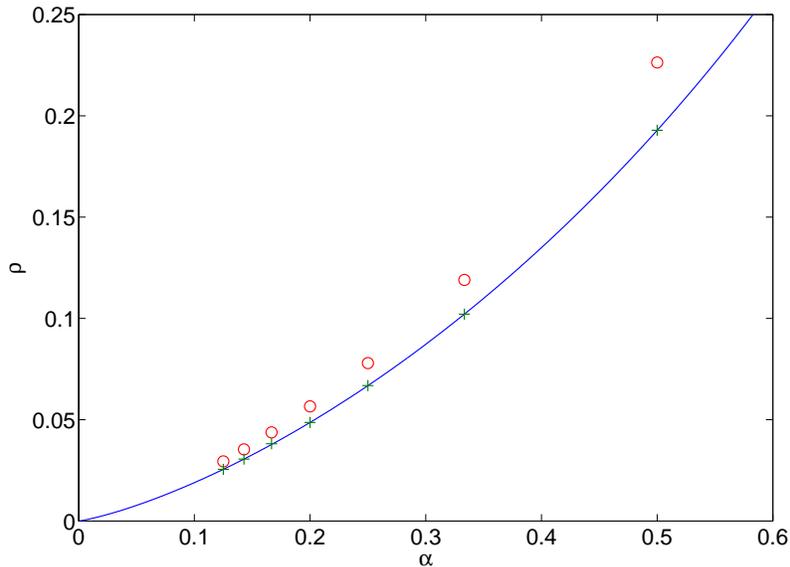}
\caption{\label{figure1}(Color online) Critical values $\rho_{\rm c}$ of total non-zero density versus compression rate $\alpha$. Circles and crosses correspond to the localized and uniform densities, respectively, for $\alpha=1/2,1/3,\ldots,1/8$. The curve represents the relation between $\rho_{\rm c}$ and $\alpha$ for the rotationally invariance matrix ensembles. Crosses coincide with values of the curve for $\alpha=1/T$ ($T=2,3,\ldots$). }
\end{figure}

However, this is not the case when the non-zero density is not uniform. As a distinctive example, we examined the case of localized density, which is characterized by setting $\rho_1=T\rho$ and $\rho_t=0$ for $t=2,3,\ldots,T$. 
\textcolor{black}{Such an assumption is plausible in handling various kinds of real world signals at least as a first approximation; 
due to the intrinsic nature of real world signals, one can expect that the density of the sparsest expression is localized 
in a certain block when the concatenation of identity and randomly chosen Fourier/wavelet bases, 
which is a representative example of the $T$-concatenation of orthonormal bases, is employed.}
Table \ref{table} and Figure \ref{figure1} show critical values of the total non-zero density $\rho=\nb^{-1}\sum_{t=1}^\nb\rho_t$ given the compression rate $\alpha=1/T$ for the uniform and localized density cases. These show that the concatenated matrices always result in better $l_1$-recovery performance for vectors of the localized densities, and the significance increases as $T$ becomes smaller while matrices of rotationally invariant ensembles result in identical performance as long as $\rho$ is unchanged. 
\textcolor{black}{It is noteworthy that the performance gain  
is obtained without utilizing the knowledge of the profile of $\rho_t$ in the recovery stage. }
This indicates that, in addition to their ease of implementation and theoretical elegance, the concatenated orthogonal matrices are preferred for practical use in terms of their high recovery performance for vectors of non-uniform non-zero densities. 
 
\begin{table}
\caption{\label{table} Comparison of critical values of total non-zero density $\rho=\nb^{-1}\sum_{t=1}^\nb\rho_t$ for uniform and localized densities. The values are rounded off to the fourth decimal. The values for uniform density is identical to those of the rotationally invariant matrix ensembles for compression rate $\alpha=M/N=1/T$.}
\begin{center}
\begin{tabular}{c|cccccccc}
$T$&2&3&4&5&6&7&8\cr
\hline 
uniform & 0.1928 & 0.1021 & 0.0668 &0.0487 & 0.0378 & 0.0308 & 0.0257 \cr
localized& 0.2267 & 0.1190 & 0.0780 & 0.0566 & 0.0438 &0.0354 & 0.0294
\end{tabular}
\end{center}
\end{table}

\subsection{Numerical justification}

\begin{figure}
	\centering
		\includegraphics[width=0.47\columnwidth]{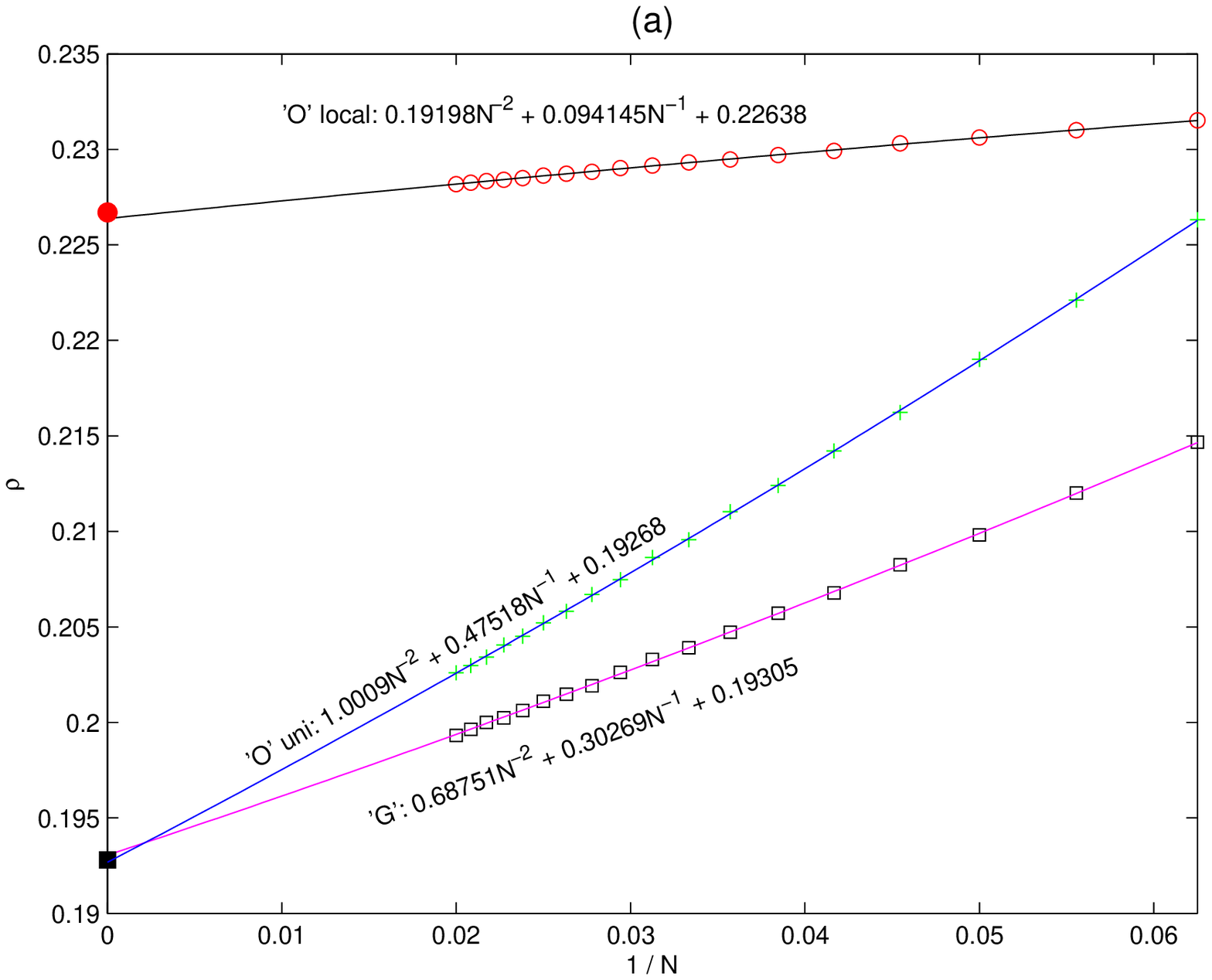}%
		\hspace*{0.05\columnwidth}
		\includegraphics[width=0.47\columnwidth]{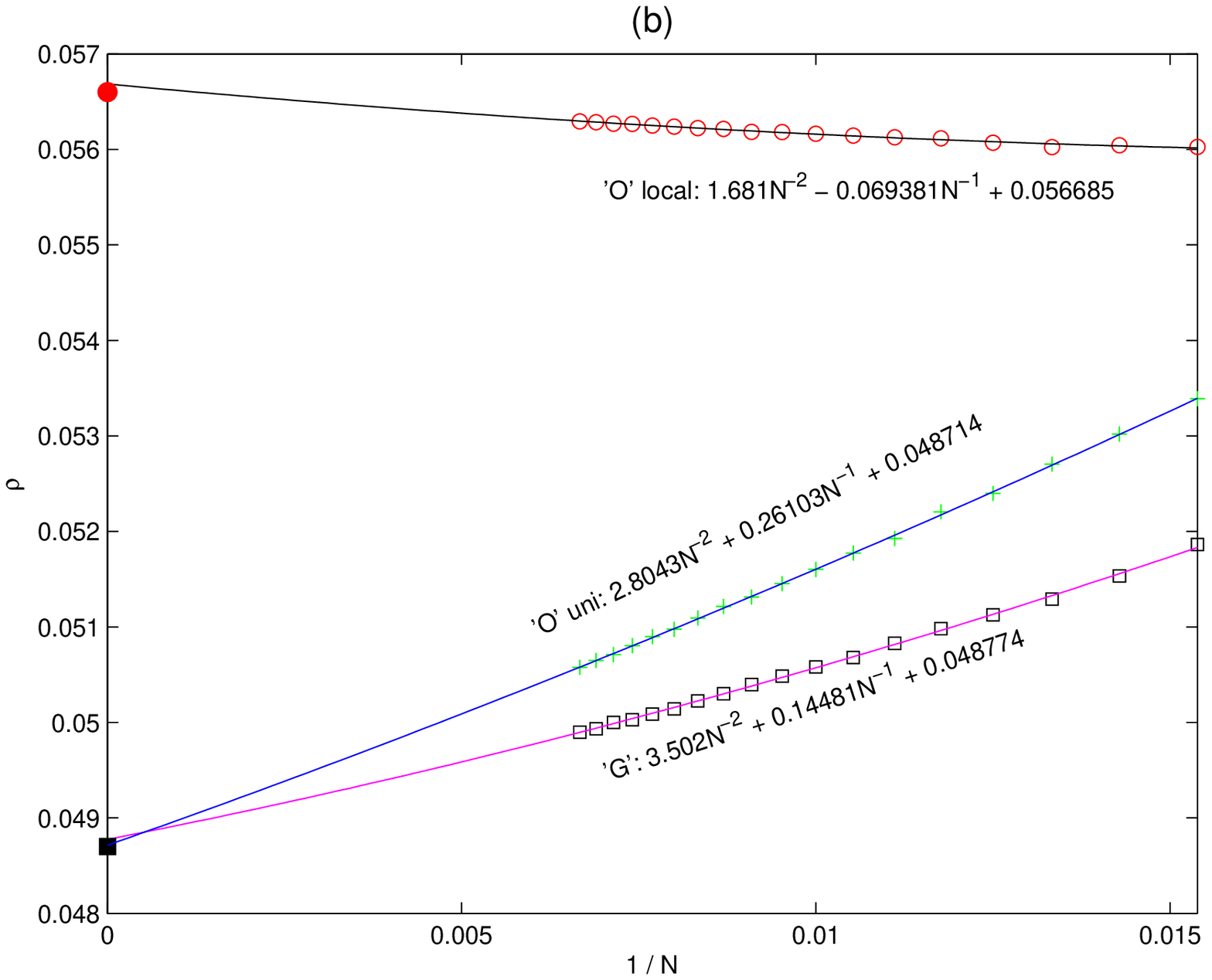}
	\caption{\label{figure2}(Color online) Experimental assessment of critical densities for $l_1$-reconstruction. Experimental data (see the main text) were fitted with a quadratic function of $1/N$ and plotted with solid lines. Concatenated orthogonal ``O'' and i.i.d. standard Gaussian ``G'' basis under uniform and localized densities. Filled markers represent the predictions obtained through the replica analysis. Extrapolation for $N = T M\to \infty$ provides the estimates for the critical values $\rho_{\rm c}$. (a)  $T = 2$, where the markers correspond to simulated values $M = 8,9,\ldots,25$, and (b) $T = 5$, where the markers correspond to simulated values $M = 13,14,\ldots,30$. }
\end{figure}

To justify our theoretical results, we conducted extensive numerical experiments of the $l_1$-reconstruction. Figures~\ref{figure2}~(a)~and~(b) depict the experimental assessment of the critical threshold for $T=2$ and $T=5$, respectively. The case of an i.i.d.\ standard Gaussian dictionary is also plotted for comparison. Given fixed values of $T$ and $N$, a trial was started with an empty vector $\vm{x}^0$ and a concatenated orthogonal dictionary generated from a set of $T$ standard i.i.d.\ $M \times M$ Gaussian matrices using QR-decomposition. Based on the relative densities $\{\rho_{t}\}$, one sub-vector $\vm{x}^0_t$ was then randomly chosen and assigned a non-zero component drawn from the standard Gaussian ensemble. Matlab algorithm ``linprog'' from Optimization Toolbox was used to solve the $l_1$-minimization problem and obtain the reconstruction $\hat{\vm{x}}$. The reconstruction was deemed to be a success if $\|\vm{x}^0 -\hat{\vm{x}}\|_{1} < 10^{-6}$ and a failure otherwise. Given a successful reconstruction, we again randomly chose one sub-vector $\vm{x}^0_t$ based on the densities $\{\rho_{t}\}$ and inserted a non-zero component drawn independently from the standard Gaussian ensemble into it. The process was continued until the original vectors $\{\vm{x}^0_t\}$ had $\{K_t\}$ non-zero components and the reconstruction was deemed a failure, that is, $\|\vm{x}^0 -\hat{\vm{x}}\|_{1} > 10^{-6}$. The critical value $K_{\mathrm{c}} = \sum_{t}K_{t} - 1$ was recorded and the experiment was started again using a new independent dictionary and an empty vector $\vm{x}^0$. For each value of $T$ and $N$, we carried out $10^6$ independent trials. The experimental critical density was defined as $\rho_{\mathrm{c}}(T,N) = \overline{K_{\mathrm{c}}} / N$, where $\overline{\cdots}$ denotes the arithmetic average over the trials. For all system sizes, we also computed the experimental per-block densities $\{\overline{K_{t}} / M\}$ and checked that they were close to the desired densities $\{\rho_{t}\}$ after the $10^6$ trials. For fixed $T$, the experimental data points $\rho_{\mathrm{c}}(T,N)$ were fitted with a quadratic function of $1/N$. Extrapolation for $N \to \infty$ provided the experimental estimates of the critical densities, as listed in Table~\ref{table2} in which the theoretical estimates in Table~\ref{table} are also listed for comparison.
\begin{table}
	\caption{\label{table2} Comparison between experimental and theoretical assessments of critical values of total non-zero density $\rho_{\mathrm{c}}$ for uniform and localized densities. The values are rounded off to the fourth decimal. For all cases in the table, the differences between the two estimates are most evident in the last digits. The excellent agreement to the experimental assessments justifies our theoretical treatment.}
		\begin{center}
			\begin{tabular}{c|ccccc}
				$T$&2&3&4&5\cr
				\hline 
				uniform (experiment) & 0.1927 & 0.1019 & 0.0670 & 0.0487  \cr
				uniform (theory) & 0.1928 & 0.1021 & 0.0668 & 0.0487 \cr
				\hline
				localized (experiment) & 0.2264 & 0.1196 & 0.0779 & 0.0567 \cr
				localized (theory) & 0.2267 & 0.1190 & 0.0780 & 0.0566
			\end{tabular}
		\end{center}
\end{table}
Comparing the theoretical and experimental results confirms the accuracy of the replica analysis. From Figure~\ref{figure2}, we observe that for the finite-sized systems, the $T$ orthogonal dictionaries seem to always provide higher thresholds $\rho_{\mathrm{c}}$ than the Gaussian one, even for uniform densities.

\section{Summary}
In summary, we investigated the performance of recovering a sparse vector $\vm{x}\in \mathbb{R}^N$ from a linear underdetermined equation $\vm{y}=\vm{D}\vm{x} \in \mathbb{R}^M$ when $\vm{D}$ is provided as a concatenation of $T=N/M$ independent samples of $M \times M$ random orthogonal matrices and the $l_1$-recovery scheme is used. Performance was measured using a threshold value of the density $\rho$ of non-zero entries in the original vector $\vm{x}^0$, below which the $l_1$-recovery is typically successful for given compression rate $\alpha=M/N=T^{-1}$. For evaluating this, we used the replica method in conjunction with the development of an integral formula for handling the random orthogonal matrices. Our analysis indicated that the threshold is identical to that of the standard setup for which matrix entries are sampled independently from identical Gaussian distribution when the non-zero entries in $\vm{x}^0$ are distributed uniformly among $T$ blocks of the concatenation. However, it was also shown that the concatenated orthogonal matrices generally provide higher threshold values than the standard setup when the non-zero entries are localized in a certain block. Results of extensive numerical experiments exhibited excellent agreement with the theoretical predictions. These mean that, in addition to their ease of implementation and theoretical elegance, the concatenated orthogonal matrices are preferred for practical use in terms of their high recovery performance for vectors of non-uniform non-zero densities.  

Promising future studies include performance evaluation in the case of noisy situations and development of approximate recovery algorithms suitable for the concatenated orthogonal matrices \textcolor{black}{\cite{VilaSchniter,Krzakala2012a,Krzakala2012b}.} 
 
\ack
The authors would like to thank Erik Aurell, Mikael Skoglund, and Lars Rasmussen for their useful comments. We also thank CSC --- IT Center for Science Ltd.\ for the allocation of computational resources. This work was partially supported by grants from the Japan Society for the Promotion of Science (KAKENHI Nos. 22300003 and 22300098) (YK) and Swedish Research Council under VR Grant 621-2011-1024 (MV). 

\appendix

\section{Derivation of (\ref{formula_concrete})}
\label{appendix1}
When $\vm{O}_t$ are sampled independently and uniformly from the Haar measure of the $M\times M$ orthogonal matrices, $\tilde{\vm{u}}_t=\vm{O}_t \vm{u}_t$  $(t=1,2,\ldots,T)$ are distributed independently and uniformly on the surfaces of the $M$-dimensional hyperspheres of radius $\sqrt{M v_t}$ for a fixed set of $M$-dimensional vectors $\vm{u}_t$ satisfying $|\vm{u}_t|^2=M v_t$. This means that the integral of (\ref{formula_def}) can be expressed as
\begin{eqnarray}
&&\frac{\int \left (\prod_{t=1}^T {\cal D}\vm{O}_t \right ) \delta \left (\sum_{t=1}^T \vm{O}_t\vm{u}_t \right ) }
{ \int \left (\prod_{t=1}^T {\cal D}\vm{O}_t \right ) } \cr
&&=\frac{\int \left (\prod_{t=1}^T d\tilde{\vm{u}}_t \delta (|\tilde{\vm{u}}_t|^2-Mv_t) \right ) 
\delta \left (\sum_{t=1}^T \tilde{\vm{u}}_t  \right )}
{\int \left (\prod_{t=1}^T d\tilde{\vm{u}}_t \delta (|\tilde{\vm{u}}_t|^2-Mv_t) \right )}. 
\label{spherical_integration}
\end{eqnarray}
We insert the Fourier expressions of $\delta$-function 
\begin{eqnarray}
\delta (|\tilde{\vm{u}}_t|^2-Mv_t) =\frac{1}{4 \pi \sqrt{-1}} \int_{-\sqrt{-1}\infty}^{+ \sqrt{-1}\infty}
d \Lambda_t \exp \left (-\frac{\Lambda_t}{2} (|\tilde{\vm{u}}_t|^2-Mv_t) \right )
\label{delta1}
\end{eqnarray}
and
\begin{eqnarray}
\delta \left (\sum_{t=1}^T \tilde{\vm{u}}_t  \right ) =\frac{1}{(2 \pi )^M} \int_{-\infty}^{+ \infty}
d \vm{k} \exp \left (\sqrt{-1} \vm{k} \cdot \left (\sum_{t=1}^T \tilde{\vm{u}}_t  \right )\right )
\label{delta2}
\end{eqnarray}
into the numerator of (\ref{spherical_integration}), and carry out the integration with respect to $\{\tilde{\vm{u}}_t\}$, where $\vm{k}=(k_1,k_2,\ldots,k_M)$. This yields the expression 
\begin{eqnarray}
&&\int \left (\prod_{t=1}^T d\tilde{\vm{u}}_t \delta (|\tilde{\vm{u}}_t|^2-Mv_t) \right ) 
\delta \left (\sum_{t=1}^T \tilde{\vm{u}}_t  \right ) \cr
&& \propto 
\int \left (\prod_{t=1}^T d\Lambda_t \Lambda_t^{-M/2}\right )  \exp\left (\sum_{t=1}^T\frac{M \Lambda_t v_t}{2} \right ) 
\int d \vm{k} \exp \left (-\frac{\left (\sum_{t=1}^T \Lambda_t^{-1} \right ) }{2} |\vm{k}|^2 \right ) \cr
&& \propto  
\int \left (\prod_{t=1}^T d\Lambda_t \Lambda_t^{-M/2}\right )  \exp\left (\sum_{t=1}^T\frac{M \Lambda_t v_t}{2} \right ) 
\left (\sum_{t=1}^T \Lambda_t^{-1} \right )^{-M/2}. 
\end{eqnarray}
Evaluating this by means of the saddle point method with respect to $\Lambda_t$ $(t=1,2,\ldots, T)$ results in
\begin{eqnarray}
&&\lim_{M \to \infty} \frac{1}{M}\ln \left (
\int \left (\prod_{t=1}^T d\tilde{\vm{u}}_t \delta (|\tilde{\vm{u}}_t|^2-Mv_t) \right ) 
\delta \left (\sum_{t=1}^T \tilde{\vm{u}}_t  \right )  
\right ) \cr
&&= \mathop{\rm extr}_{\Lambda_1,\Lambda_2,\ldots,\Lambda_T}
\left \{ -\frac{1}{2}\ln \left (\sum_{t=1}^T \Lambda_t^{-1} \right )
-\frac{1}{2}\sum_{t=1}^T\ln (\Lambda_t)+\frac{1}{2}\sum_{t=1}^T \Lambda_t v_t \right \}. 
\label{numer}
\end{eqnarray}
Similarly, the denominator of (\ref{spherical_integration}) is evaluated as
\begin{eqnarray}
\lim_{M \to \infty} \frac{1}{M} \ln \left (
\int \left (\prod_{t=1}^T d\tilde{\vm{u}}_t \delta (|\tilde{\vm{u}}_t|^2-Mv_t) \right ) \right )
=\frac{1}{2}\sum_{t=1}^T \ln (v_t)+\frac{T}{2}.
\label{denomi}
\end{eqnarray}
Substituting (\ref{spherical_integration}), (\ref{numer}), and (\ref{denomi}) into (\ref{formula_def}) leads to the expression of (\ref{formula_concrete}). 

\section*{References}
\providecommand{\newblock}{}

\end{document}